\documentclass[twocolumn,superscriptaddress,10pt]{revtex4-2}
\usepackage{gnuplottex}
\usepackage{mathtools}
\usepackage{float}
\usepackage{array}
\usepackage[english]{babel}
\usepackage[utf8]{inputenc}
\usepackage{amsmath}
\usepackage[table]{xcolor}
\usepackage{amsfonts}
\usepackage{graphicx}
\usepackage{dsfont}
\usepackage[colorlinks=true, linkcolor=blue, citecolor=dred]{hyperref}
\usepackage{algorithm}
\usepackage[noend]{algpseudocode}
\usepackage{bbold}
\usepackage{mathtools}
\usepackage{amssymb}
\usepackage{bm}
\usepackage{bbm}
\usepackage{tabularx, booktabs}
\usepackage{xspace}
\usepackage{listings}
\usepackage[noend]{algpseudocode}
\newcolumntype{Y}{>{\centering\arraybackslash}X}
\definecolor{dgreen}{rgb}{0,.5,0}
\definecolor{dblue}{rgb}{0,0,.5}
\definecolor{dred}{rgb}{0.5,0,.5}

\usepackage{physics}

\begin{document}

 \title{
 % The nonlinear quantum dimer in a cavity : genuine quantum self-trapping\\
  % {Polaritonic Energy Transfer in an Anharmonic Vibrational Dimer: \\
  % From Enhanced Self-Trapping to Enhanced Energy Transport}
% Light-Induced Self-Trapping and Enhanced Transport in a Cavity-Controlled Anharmonic Vibrational Dimer
% Polaritonic Enhancement of Self-Trapping and Transition to Energy Transport in an Anharmonic Vibrational Dimer
 % \saad{Polaritonic Control of Energy Transport in an Anharmonic Vibrational Dimer: \\Emergence of Genuine Cavity-Induced Quantum Self-Trapping  }
 % From Enhanced Self-Trapping to Energy Transport in a Cavity-Controlled Vibrational Dimer
 % From Enhanced Self-Trapping to Energy Transport in a Cavity-Controlled Vibrational Dimer
%  Cavity-Controlled Energy Transfer in an Anharmonic Quantum Dimer:\\ 
% From Light-Induced Self-Trapping to Enhanced Transport\\
Light-Induced Quantum Self-Trapping of Vibrational Excitons in an Optical Cavity
 } 
 
 %from photon-enhanced self-trapping to cavity-assisted energy flow}

\author{Vincent Pouthier}
% \email{vincent.pouthier@univ-fcomte.fr }
\affiliation{Universit\'{e} Marie et Louis Pasteur, CNRS, Institut UTINAM (UMR 6213), Equipe $\phi$th, F-25000 Besan\c {c}on, France} 

\author{Saad Yalouz} 
\email{saad.yalouz@cnrs.fr}
\affiliation{Laboratoire de Chimie Quantique de Strasbourg, Institut de Chimie,
CNRS/Université de Strasbourg, 4 rue Blaise Pascal, 67000 Strasbourg, France}

\begin{abstract}

 In an optical cavity, strong light--matter coupling between excitons and photons has been widely reported as a way to enhance energy delocalization through spatially extended polaritonic states.
In contrast, leveraging cavity-mediated light--matter effects to promote the reciprocal phenomenon, namely \textit{energy localization}, remains largely underexplored.
In the present  
work, we address this question by focusing on a special form of energy localization arising from nonlinear  
matter interactions: \textit{Quantum Self-Trapping} (QST).
We employ a generalized Tavis--Cummings model to investigate the transport of vibrational excitons---\textit{i.e., vibrons}---between two anharmonic vibrational modes and examine their interplay with cavity photons.
In the absence of a cavity, the arising of true and complete QST---\textit{i.e.}, an infinite-lifetime 
localization---is not possible due to the symmetry of the system. 
The energy transfer between the two modes still occurs, slowed down by the many-body interactions.
Coupling the system to a single-mode cavity strongly alters this behavior, with two emerging regimes.
First, at weak light--matter coupling, destructive interference between newly opened transition pathways suppresses energy exchange, leading to cavity-enhanced self-trapping.
As the coupling strength increases, these interference effects evolve leading to cavity-assisted energy transfer, where we observe an acceleration of the vibrational energy flow.
Most notably, we identify critical coupling strengths that separate both regimes in which the dynamics almost totally freeze, suggesting the arising of a ``stabilized'' light-induced~QST of many-vibron bound states.
These results suggest that optical cavities can not only enhance transport but could also stabilize energy localization phenomena, providing a new route to control energy flow in quantum systems.

\end{abstract}

\maketitle

\section{Introduction}

% Polaritonic chemistry and physics explore the novel behaviors that arise when molecules or materials couple to the confined electromagnetic field in a cavity. 
% Under such conditions, hybrid light-matter quantum states, known as polaritons, form as a coherent superpositions of photonic and matter excitations.
% These exotic states have been observed across a wide range of platforms, including electron spins~\cite{kubo10}, inorganic crystals~\cite{shelton11,luxmoore14,sivarajah19}, cold atoms~\cite{kaluzny83,raizen89,thompson92}, and semiconductors~\cite{weisbuch92}, among others. As a central feature of polaritonic systems lies the possibility to engineer light-matter interactions by modifying the cavity environment.
Polaritonic chemistry and physics explore the novel behaviors that arise when molecules or materials couple to the confined electromagnetic field in a cavity. \linebreak
Under such conditions, hybrid light-matter quantum states, known as polaritons, form as coherent superpositions of photonic and matter excitations. 
These exotic states, which exhibit properties distinct from their uncoupled constituents, have been observed across a wide range of platforms, including electron spins~\cite{kubo10}, inorganic crystals~\cite{shelton11,luxmoore14,sivarajah19}, cold atoms~\cite{kaluzny83,raizen89,thompson92}, and semiconductors~\cite{weisbuch92,lidzey1998strong,lidzey2000photon}. 
% Crucially, as polaritonic states arise from the interplay between light and matter, their properties can be directly tuned by the cavity environment which conditions the strength and character of these interactions.
% Crucially, because polaritonic states naturally depend on the strength and character of light–matter coupling, their properties can be directly tuned by modifying the cavity environment.
Crucially,  polaritonic states properties can be directly tuned by the cavity environment which conditions the strength and character of the light-matter interactions.
This tunability has sparked growing interest in exploiting the resulting cavity-mediated light-matter effects to influence a variety of physicochemical processes. 
Examples include chemical reactivity~\cite{herrera16,groenhof18}, molecular emission~\cite{wang14}, and excited-state dynamics~\cite{fabri2020lici-a,fabri2020lici-b,csehi2017lici,fabri2022probing}, among others.

% Within the broad landscape of processes that can be modulated via cavity-mediated light-matter interactions, the ctronol of quantum energy t  stands out as one of the most promising directions.
Within the broad range of applications enabled by light-matter interactions in cavity, the control of energy transport emerges as one of the most promising directions.
In this framework, numerous theoretical and experimental studies have demonstrated that energy transfer processes can be modulated \textit{via} a strong coupling between  energy carriers---\textit{i.e.}, excitons---and cavity modes (see Refs. \cite{sandik2025cavity, bhuyan2023rise} and references therein). 
% In this framework, numerous experimental works have already demonstrated the possibility to strongly couple energy carriers---\textit{i.e.}, excitons---to cavity modes in order to  (see Refs. \cite{sandik2025cavity, bhuyan2023rise} and references therein). x
As a result, a wide range of cavity-enhanced phenomena have been reported in the literature, spanning from cavity-induced excitonic delocalization to enhanced energy-transfer processes~\cite{feist2015extraordinary, hagenmuller2025disorder, schachenmayer15, coles2014strong, spano2015optical, rozenman18, hou2020ultralong, balasubrahmaniyam2023enhanced}.
These effects are enabled by the formation of spatially delocalized polaritonic states, which allow excitonic density to extend over a larger number of matter subunits compared to what is known outside a cavity.

Interestingly, while most works have focused on identifying conditions that promote cavity-enhanced energy delocalization, the reciprocal phenomenon---\textit{i.e.}, the enhancement of energy localization by a cavity---remains largely unexplored~\cite{denning2022cavity,chestnov2018heat,rokaj2023cavity}.
Addressing this question constitutes one of the central motivations of the present work.
Here, we specifically investigate a form of localization that stems from nonlinear (i.e., many-body) matter interactions, known as \textit{self-trapping}, whose physical origin is detailed below.

% \saad{Je n'ai trouve que 3 travaux abordant le sujet. Parmi lesquels une lettre (Ref.~\cite{denning2022cavity}) proche de ce que nous faisons ou les auteurs considerent aussi de l'interaction exciton-exciton dans un semi-conducteur et mettent en evidence l'idee de brisure de symmetrie induite par cavite generant de la localisation!}

Self-trapping processes have been extensively studied in a variety of media (in the absence of a cavity) for several decades~\cite{sato06, flach98, flach08}.
One of the most notable contributions to this field was made by Davydov in the 1970s, in his description of high-frequency amide-I vibrons in $\alpha$-helices~\cite{davydov73}.
Within a semiclassical continuum approximation, he modeled the vibron dynamics using a nonlinear Schrödinger equation, whose solutions give rise to the so-called Davydov solitons~\cite{scott92}.
In the mid-1980s, lattice effects were incorporated through the study of discrete versions of this nonlinear equation. These works revealed a remarkable phenomenon known as the \textit{self-trapping} mechanism~\cite{eilbeck85}, whereby energy initially localized at a given site remains confined to it over time.
Subsequently, Sievers and Takeno showed that self-trapping can be understood as a particular instance of a broader class of excitations known as discrete breathers~\cite{sievers88}, \textit{i.e.}, time-periodic, spatially localized modes arising from the interplay between discreteness and nonlinearity~\cite{flach98, flach08}.
 
At the quantum level, a direct analogue of the discrete nonlinear Schrödinger equation is provided by the attractive Bose-Hubbard model, which is widely used to describe interacting vibrational excitons--\textit{i.e., vibrons}--in molecular lattices~\cite{kimball81, persson92, falvo18, pouthierPRE03}.  
Vibron--vibron interactions give rise to many-body bound states, which constitute the quantum analogues of discrete breathers~\cite{pouthierPRE03, bogani90, wright93, scott90, scott94, dorignac04, proville05, falvo06}. In these bound states, several vibrons remain spatially localized, resulting in an energy lower than that of spatially separated excitations.  
These condensed excitations behave as a single \textit{effective} particle, capable of moving across the lattice with a well-defined momentum.  
Unlike classical discrete breathers, these quantum bound states do not yield permanent energy localization, as they preserve translational symmetry. 
The resulting \textit{Quantum Self-Trapping} process (QST)  is therefore ``unstable'' as the tunneling of the vibron condensate between lattice sites can still occur, but over very long timescales.
This transient energy localization occurs with a time that increases with both the number of vibrons and the strength of the nonlinearity (i.e., many-body interaction). 

In light of these considerations, it is clear that, when considering a translationarry invariant lattice in vacuum (i.e., without a cavity), the emergence of a {genuine} stable QST—i.e., an infinite-lifetime, many-body–driven energy localization—can occur only if lattice symmetry is broken (see Refs.~\cite{pouthierPRB07, pintoPRA09, pinto09, pouthierPRE22, pouthierJCP22} for illustrations).
In the present work, we will explore an alternative route: embedding a lattice in an optical cavity. 

To this end, we consider a model system, namely an anharmonic quantum dimer (AQD), which is known to exhibit “unstable” QST in vacuum due to its symmetry~\cite{scott90,bernstein93,pouthierphysicaD06}. Using a Tavis–Cummings model~\cite{tavis68,tavis69}, we describe the coupling of the dimer to a cavity mode and investigate how light–matter interactions influence energy transport between its two local anharmonic vibrational modes~\cite{hernandez19,tao22,javier15}. We show that the hybridization between vibrons and confined photons gives rise to rich dynamical behavior, ranging from cavity-enhanced self-trapping to cavity-assisted energy transfer. In particular, we identify critical light–matter coupling strengths that separate these regimes and lead to a “stabilized” light-induced QST, in which many-vibron bound states become fully localized.

The paper is organized as follows.
In Sec.~\ref{sec:Theo}, the AQD embedded in an optical cavity is introduced with the associated light--matter Hamiltonian.
Then the relevant observables required for characterizing the dynamics are described. 
In Sec.~\ref{sec:Num}, a numerical analysis is performed for characterizing energy exchanges between the two modes of the dimer. Then, in Sec.~\ref{sec:discussion}, the results are discussed and interpreted using a phenomenological four-level model.
We close the paper with concluding remarks in Sec.~\ref{sec:conclu}.

\section{Theoretical background} 
\label{sec:Theo}

\subsection{Light-Matter Hamiltonian }

\begin{figure}[t]
    \centering
    \includegraphics[width=7.5cm]{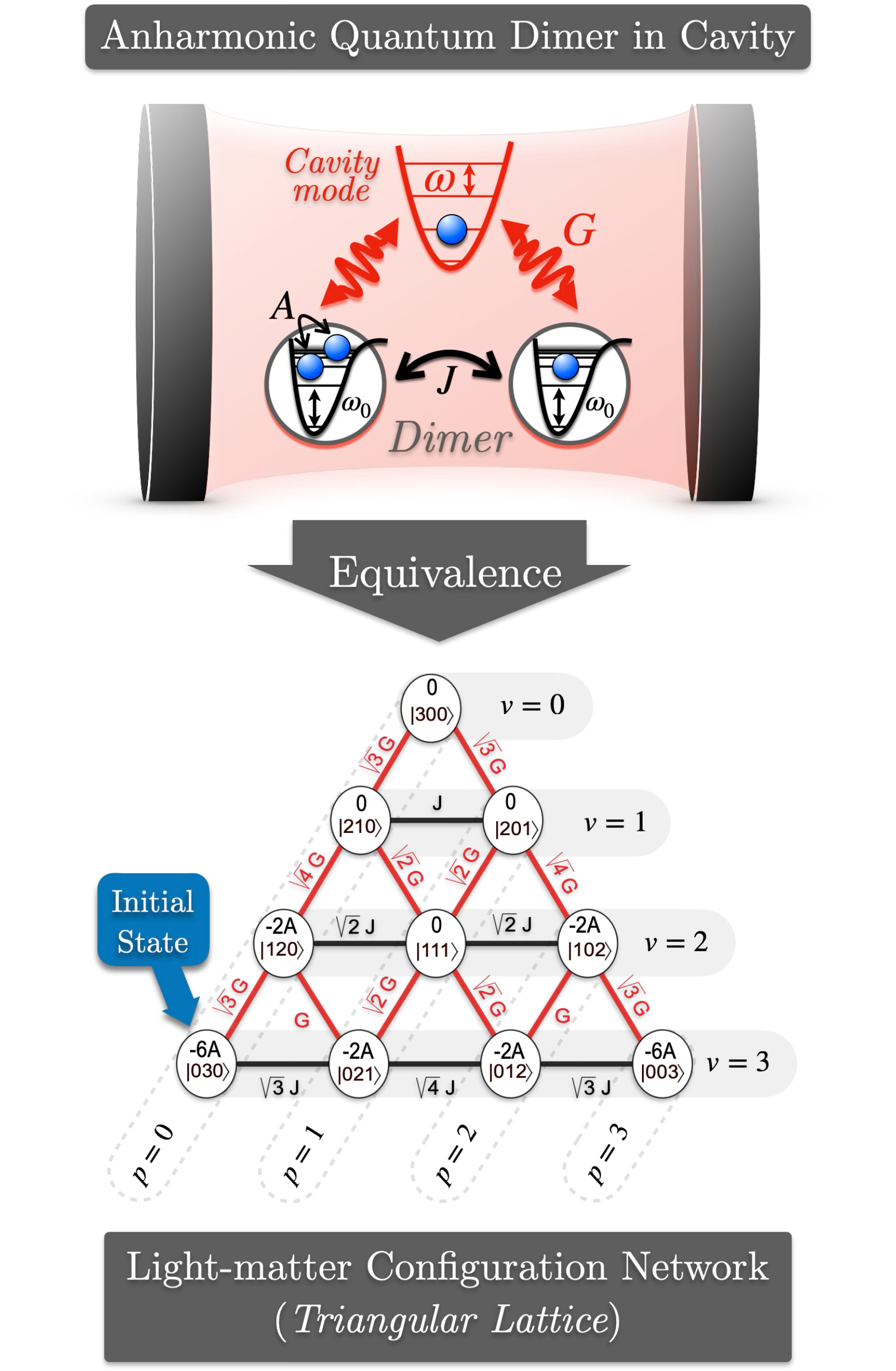}
    \caption{ 
    Anharmonic Quantum Dimer  in an optical cavity. \linebreak
  \textit{Top panel:} Physical system consisting of two local anharmonic vibrational modes with frequency $\omega_0$, coupled via dipole--dipole interaction $J$ and on-site many-body interaction $A$, and interacting with a cavity mode of frequency $\omega$ and coupling strength $G$. 
\textit{Bottom panel:} When embedded in a cavity, the matrix representation of the light--matter Hamiltonian $\mathcal{H}$ (see Eq.~(\ref{eq:LM_Ham}) and Eq.~(\ref{eq:EJG1})) maps onto a triangular lattice.
The mapping is shown for total boson number $N_B=3$. 
Lattice sites labeled by $(v,p)$ correspond to number states $|N_B - v,\; v - p,\; p\rangle$, each characterized by an energy $E_{vp}^{(N_B)}$.
Horizontal nearest-neighbor couplings are given by $J_p^{(v)}$ (vibrational interaction), while vertical ones are given by $G_p^{(v)}$ (light--matter coupling), as defined in Eq.~(\ref{eq:EJG}).
    }
    \label{fig:networks}
\end{figure}

As illustrated in Figure~\ref{fig:networks}, we consider in our study an AQD %Anharmonic Quantum Dimer (AQD) 
composed of two coupled vibrational modes.  
Following Kimball \textit{et al.}~\cite{kimball81}, the anharmonic nature of these modes is taken into account using a Bose version of the two-site Hubbard model as described by the Hamiltonian (assuming $\hbar = 1$)
\begin{equation} \label{eq:AQD_H}
    H=\sum_{x=1}^{2} \left( \omega_0 \ b_x^{\dag}b_x-A \ b_x^{\dag}b_x^{\dag}b_x b_x \right)+J(b_1^{\dag}b_2+b_2^{\dag}b_1). 
\end{equation}
Here, $b_x^{\dagger}$ and $b_x$ are bosonic creation and annihilation operators of vibrons associated with the $x$-th local mode (with $x=1$ or $2$).  
The parameter $J$ represents the vibron hopping constant, which arises from the dipole--dipole coupling between the two modes.  
The mode frequency $\omega_0$ corresponds to the energy of a single vibron, whereas the anharmonicity $A$ characterizes the many-body attractive interaction between two vibrons located on the same mode. %local mode.

Considering now the light-matter interaction, we assume that each vibrational local mode of the AQD is coupled to the same single cavity mode with frequency $\omega$ as described by a Tavis-Cumming Hamiltonian
\begin{equation}\label{eq:LM_Ham}
    \mathcal{H}=H+\omega a^{\dag}a+\sum_{x=1}^2 G(a^{\dag}b_x+b_x^{\dag}a),
\end{equation}
where $a^{\dag}$ and $a$ represent the photonic creation/annihilation operators of the cavity mode, and $G$ defines the strength of the light-matter coupling.
This coupling is responsible for the exchange of energy quanta between cavity and vibrational modes. Consequently, the total boson number operator $\hat{N}_B = a^{\dagger} a + b_1^{\dagger} b_1 + b_2^{\dagger} b_2$ becomes a conserved quantity of the full light--matter system.
Within this context, we can describe the light--matter configuration of the full system by employing an orthogonal basis of vibron--photon number states defined by
\begin{equation}
  |N_B-v,v-p,p\rangle = \frac{a^{\dag (N_B-v)}b_1^{\dag (v-p)} b_2^{\dag p}}{\sqrt{(N_B-v)!(v-p)!p!}} |\oslash\rangle  .
\end{equation}
Such number state $|N_B-v,v-p,p\rangle$ describes $N_B-v$ photons in the cavity mode, $v-p$ vibrons on mode $x=1$ and $p$ vibrons on mode $x=2$ (with with $v=0,...,N_B$  and $p=0,...,v$).
Within such basis, the matrix elements $\mathcal{H}^{(vv')}_{pp'}$  of the light--matter Hamiltonian defined in Eq.~(\ref{eq:LM_Ham}) are given by
\begin{eqnarray}\label{eq:EJG1}
  \mathcal{H}^{(vv')}_{pp'}&=& E^{(v)}_p \delta_{p',p}\delta_{v',v}  \\
  &+&J^{(v)}_p \delta_{p',p+1}\delta_{v',v}+J^{(v)}_{p'} \delta_{p',p-1}\delta_{v',v} \nonumber \\
  &+&G^{(v)}_p \delta_{p',p}\delta_{v',v+1}+G^{(v')}_{p'} \delta_{p',p}\delta_{v',v-1} \nonumber \\
  &+&G^{(v)}_{v-p} \delta_{p',p+1}\delta_{v',v+1}+G^{(v')}_{v'-p'} \delta_{p',p-1}\delta_{v',v-1} \nonumber
\end{eqnarray}
where 
\begin{eqnarray}\label{eq:EJG}
 E^{(v)}_p&=&N_B \omega +v(\omega_0-\omega)-v(v-1)A+2p(v-p)A \nonumber \\
 J^{(v)}_p&=&\sqrt{(v-p)(p+1)} J\nonumber \\
 G^{(v)}_p&=&\sqrt{(N_B-v)(v-p+1)} G
\end{eqnarray}
Interestingly, as illustrated in the bottom panel of Fig.~\ref{fig:networks}, the dynamics of the light--matter system is isomorphic to that of an \textit{effective} single particle moving on a triangular lattice (represented here for $N_B=3$ bosons). 
Each effective site represents a specific light--matter configuration characterized by an energy $E_p^{(v)}$ and labeled by a pair of indices $(v,p)$, with $v=0,1,\dots,N_B$ and $p=0,1,\dots,v$.
Structurally, the triangular lattice can be viewed as a set of horizontal chains of increasing length (from top to bottom). Each chain of size $v=1,\dots,N_B$ describes the pure vibrational dynamics of the dimer within the $v$-vibron subspace. Horizontal black links between neighboring sites $(v,p)$ and $(v,p+1)$ are described by the hopping amplitude $J_p^{(v)}$.
Due to the presence of light--matter coupling, these chains are not independent but are connected by vertical red links that account for energy exchange between the cavity mode and the vibrational modes. 
Couplings between $(v,p)$ and $(v+1,p)$ are described by $G_p^{(v)}$, while couplings between $(v,p)$ and $(v+1,p+1)$ are given by $G_{v-p}^{(v)}$.

Importantly, the triangular network representation shown in Fig.~\ref{fig:networks} provides a convenient framework to assess the evolution of a QST phenomenon as the light--matter coupling is varied. 
To this end, we consider the initial state $\ket{\Psi_{\mathrm{ini}}} = \ket{0, N_B, 0}$, where $N_B$ vibrons are localized in the first anharmonic mode, while the cavity contains no photons. Within this representation, this state corresponds to initializing the dynamics at the bottom-left effective site, as illustrated in Fig.~\ref{fig:networks} (see blue arrow).
Tracking the ensuing quantum dynamics from this site, and how the associated occupation probabilities evolve across the triangular network, allows us to identify the key pathways and intermediate light--matter configurations that govern vibrational energy transfer.

\subsection{Quantum Dynamics}\label{sec:QD}

To investigate the vibron--photon dynamics, we simulate the time evolution of the initial state $\ket{\Psi_{\mathrm{ini}}} = \ket{0, N_B, 0}$ according to
\begin{equation} \label{eq:psit}
    |\Psi(t)\rangle = U(t)\,|\Psi_{\mathrm{ini}}\rangle,
\end{equation}
where the time-evolution operator $U(t) = \exp(-i\mathcal{H}t)$ is constructed numerically via exact diagonalization of the Hamiltonian $\mathcal{H}$, yielding
\begin{equation}\label{eq:U}
    U(t) = \sum_{\mu} e^{-i E_{\mu} t} \, |\chi_{\mu}\rangle \langle \chi_{\mu}|.
\end{equation}
Here, $|\chi_{\mu}\rangle$ and $E_{\mu}$ denote the polaritonic eigenstates and their corresponding eigenenergies, respectively.
Based on  $|\Psi(t) \rangle$, we can then  characterize the vibrational energy transfer by computing quantum probabilities over time. 
The quantum probability  of observing the system  in a light-matter configuration $| N_B-v,v-p,p \rangle$ at time $t$ is defined as 
\begin{equation}
  \pi_{v,p}^{(N_B)}(t) =|\langle N_B-v,v-p,p|\Psi(t)\rangle|^2
\end{equation}
When one considers the isomorphism with the triangular lattice shown in Fig.~\ref{fig:networks},
such a quantum probability will define the probability for the system to occupy   the effective site $(v,p)$ at time $t$. 
Studying its spatio-temporal evolution on the triangular lattice makes it possible to track the motion of the effective single-particle on this network and determine whether it propagates or localizes. 
% This allows us to clearly identify the self-trapping phenomenon.
%
Note that in a finite-size network, the probability does not converge to a stationary value because of the unitary nature of the dynamics. Instead, it fluctuates around a long time average distribution called the limiting  probability $\bar \pi_{v,p}^{(N_B)}$. It is defined as
\begin{eqnarray}
\label{eq:limit_prob}
\bar \pi_{v,p}^{(N_B)}&=&\sum_{\mu,\mu'}\langle N_B-v,v-p,p|\chi_{\mu}\rangle \langle \chi_\mu |\Psi(0)\rangle  \\  
&\times&
\langle \Psi(0)|\chi_{\mu'}\rangle \langle \chi_{\mu'} |N_B-v,v-p,p\rangle \nonumber
\delta_{E_\mu,E_{\mu'}}
\end{eqnarray}
where $\delta_{E_\mu,E_{\mu'}}=1$ if $E_\mu=E_{\mu'}$ and zero otherwise. The limiting probability gives a good estimate of the time-dependent probability and is a key quantity to prove the localized nature of the dynamics~\cite{mulken06}.

Beyond the analysis of quantum probabilities, we also focus on a quantity that captures the %many-body
nature of vibron energy transfer and highlights the emergence of QST: the vibron population imbalance $\Delta P(t) $, defined as
\begin{equation}
\Delta P(t) = P_1(t) - P_2(t),
\end{equation}
where $P_x(t)$ denotes the vibron population on the $x$th mode at time $t$, i.e.,
\begin{equation}
P_x(t) = \langle \Psi(t) | b_x^{\dagger} b_x | \Psi(t) \rangle.
\end{equation}
This quantity provides a direct estimate of the distribution of vibrons across the two modes: $\Delta P(t) = N_B$ indicates that all $N_B$ vibrons are on the first mode, while $\Delta P(t) = -N_B$ indicates that all $N_B$ vibrons occupy the second mode.
%

% To conclude this section, we return to Eq.~(\ref{eq:psit}), whose solution requires specifying an initial quantum state. To investigate quantum self-trapping, we focus on a localized initial condition corresponding to the creation of $N_B$ vibrons on the first mode, with no photons in the cavity: $|\Psi(0)\rangle = |0,N_B,0\rangle$. 

\begin{figure}
\centering
    \includegraphics[width=8cm]{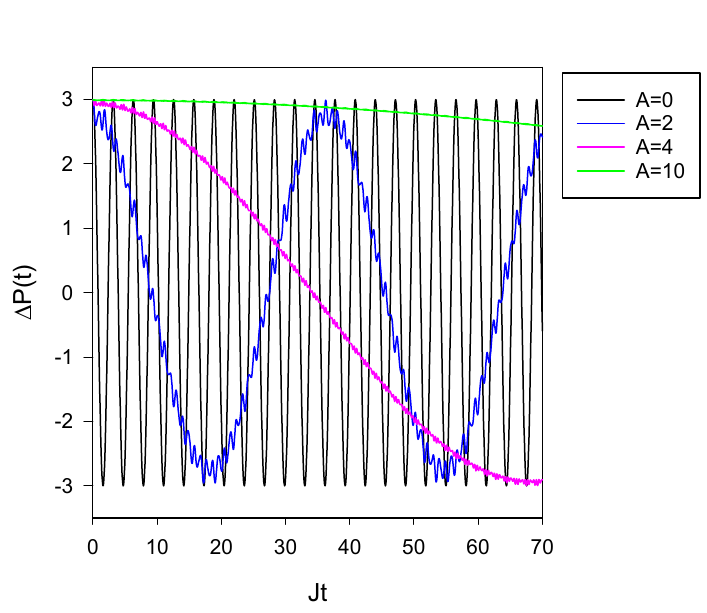}
    \caption{Illustration of the quantum self-trapping phenomenon for $N_B=3$, $J=1$, $\omega=\omega_0$ and $G=0$. The figure shows the time evolution of the vibron population imbalance for $A=0$ (black curve) and $A=4$ (red curve).}
\label{fig:A0vsA4}
\end{figure}

\section{Numerical Results} 
\label{sec:Num}

\begin{figure*}
    \centering
    \includegraphics[width=17cm]{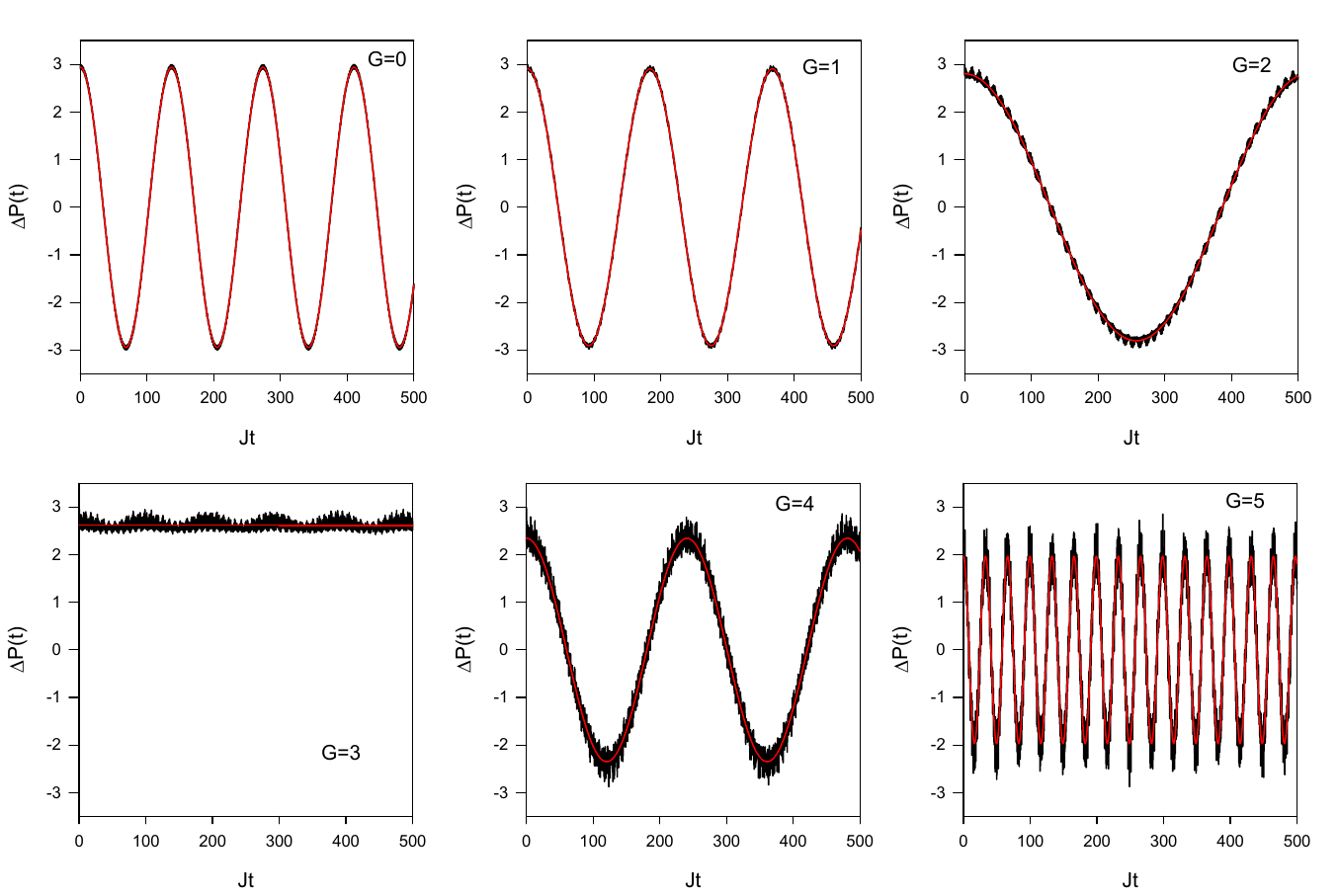}
    \caption{$G$ dependence of the time evolution of the vibron population imbalance for $N_B=3$, $J=1$, $A=4$ and $\omega=\omega_0$. The red curve defines the smoothed vibron population imbalance. It is numerically constructed by restricting the expansion of the evolution operator over the two lowest energy levels (see Eq.(\ref{eq:U})).}
\label{fig:DP}
\end{figure*}

In this section, we numerically illustrate the different dynamical behaviors that arise when the AQD is coupled to a single cavity mode.
We assume that the cavity is initially in the vacuum state, while all %bosons are condensed 
vibrons occupy the first local anharmonic vibrational mode, \textit{i.e.}, $\ket{\Psi_{ini}} = \ket{0, N_B, 0}$. 
Furthermore, we will assume a resonance between the cavity and the vibrational modes, i.e. $\omega = \omega_0$. 
In all our simulations, we take the hopping constant as the energy unit, \textit{i.e.}, $J = 1$.
% Furthermore, we assume that the cavity mode is resonant with the vibrational modes, $\omega = \omega_0$, and that the cavity is initially in the vacuum state, while all bosons are condensed in the first local anharmonic vibrational mode, \textit{i.e.}, $\ket{\Psi(0)} = \ket{0, N_B, 0}$.
% The total number of bosons will be fixed to $N_B=3$.

 \subsection{No Light--Matter Coupling $G=0$: \\ ``Unstable'' Vibronic QST }

In a first step, we consider the case without cavity coupling and briefly recall known results on the existence of ``unstable'' vibronic QST. In the absence of light--matter interaction, energy transfer in the AQD corresponds to a ``cavity-free'' process, which can be rationalized following Refs.~\cite{bernstein93,pouthierphysicaD06}. When $A\gg J$, the quantum dynamics of the system is effectively described by two low-lying bound states arising from the hybridization of the configurations $|0, N_B, 0\rangle$ and $|0, 0, N_B\rangle$, which correspond to situations where all vibrons are localized on either the left or the right site of the vibrational dimer. As illustrated in the bottom part of Fig.~\ref{fig:networks}, these two configurations occupy the left and right extremities of the lowest horizontal chain of the triangular lattice.
These states do not interact directly, but instead couple indirectly via intermediate many-vibron configurations that preserve the total number of vibrons, $v = N_B$. Within this framework, one can estimate an effective coupling between the fully localized states $|0, N_B, 0\rangle$ and $|0, 0, N_B\rangle$ given by
\begin{equation}\label{eq:Jeff}
J_{\mathrm{eff}}(N_B) = \frac{(-1)^{N_B-1} N_B J^{N_B}}{(N_B-1)! (2A)^{N_B-1}},
\end{equation}
which describes the Rabbi oscillation of the $N_B$-vibron wave packet between the two modes of the AQD~\cite{bernstein93,pouthierphysicaD06}. As indicated by Eq.~(\ref{eq:Jeff}), $|J_{\mathrm{eff}}|$ decreases rapidly as both the anharmonicity $A$ and the vibron number $N_B$ increase. Consequently, when $N_B$ vibrons are initially created on the first local vibrational mode, they remain quasi-localized on this site over a timescale that grows with $A$ and $N_B$.

To illustrate this phenomenon, Fig.~\ref{fig:A0vsA4} shows the time evolution of the vibron population imbalance for different values of the many-body interaction $A$ and for $N_B=3$. When $A=0$ , the AQD Hamiltonian $H$ (see Eq.(\ref{eq:AQD_H})) is diagonalized by introducing bright and dark normal modes $b_{\pm}=(b_1\pm b_2)/\sqrt{2}$ with eigenfrequency $\omega_{\pm}=\omega_0\pm J$. Consequently, a fast and perfect energy transfer takes place between the two modes according to a timescale specified by the Bohr frequency $2J$. Initially localized on the mode $x=1$, the vibrons reach the mode $x=2$ at the transfer time $\tau=\pi/2J\approx1.57$. When $A\neq0$, a different behavior arises. Although the energy does not localize, it takes an extremely long time to propagate from one mode to the other. The dynamics is now governed by the effective hopping constant $J_{\mathrm{eff}}(3)$ which is equal to $9.37\times10^{-2}$, $2.34\times 10^{-2}$ and $3.75\times 10^{-3}$ for $A=2$, $4$ and $10$, respectively. The transfer time $\tau=\pi/2J_{\mathrm{eff}}(3) $  increases with $A$ and reaches $16.75$ for $A=2$ and $67.02$ for $A=4$. For $A=10$, it is equal $418.87$ and becomes larger than the timescale considered in the figure. 

Importantly, Fig.~\ref{fig:A0vsA4} shows that although localization occurs, the resulting QST remains inherently “unstable”, as the vibron population eventually spreads at long times.

%Importantly, we emphasize that although localization is present, the resulting QST is inherently ``unstable'': at long times, the vibron population will eventually propagate.

 \subsection{Presence of Light--Matter Coupling $G>0$:\\  From Light-Induced QST to Enhanced Transport}

When the dimer interacts with cavity photons, new dynamical regimes emerge depending on the strength of the light–matter coupling.
We will illustrate these regimes and highlight associated exotic features, in particular the emergence of a ``stabilized'' light-induced QST.

 \subsubsection{Population Imbalance $\Delta P$: \\ Evidence of Different Dynamical Regimes }
 
Fig.~\ref{fig:DP} shows the time evolution of the vibron population imbalance $\Delta P(t)$ for $N_B=3$, $A=4$, and increasing values of light-matter coupling $G=0 \rightarrow5$. For $G=0$, it illustrates the ``unstable'' pure vibronic QST phenomenon previously displayed in Fig.~\ref{fig:A0vsA4}, but over a longer timescale.
One thus clearly distinguishes the low frequency oscillations of the vibron population imbalance, which reflect the delocalization of the vibrons between the two states $|0,3,0\rangle$ and $|0,0,3\rangle$. 
\begin{figure}
%    \centering
    \includegraphics[width=7cm]{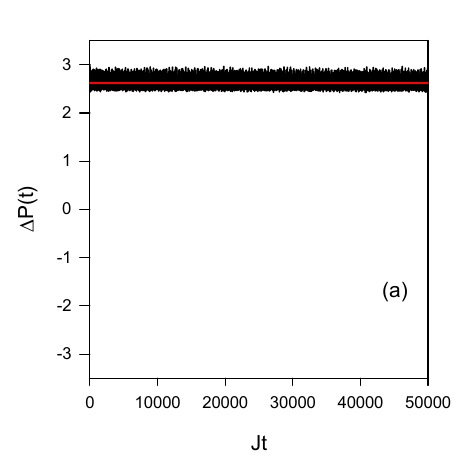}
    \caption{System dynamics at the critical point $G_c=$ for $N_B=3$, $J=1$, $A=4$ and $\omega=\omega_0$. Here we illustrate the time evolution of the vibron population imbalance $\Delta P(t)$. }
\label{fig:COURBEGC}
\end{figure}
%
% Indeed, when $A=0$ (not drawn), the AQD Hamiltonian $H$ (see Eq.(\ref{eq:AQD_H})) is diagonalized by introducing bright and dark normal modes $b_{\pm}=(b_1\pm b_2)/\sqrt{2}$ with eigenfrequency $\omega_{\pm}=\omega_0\pm J$. Consequently, a fast and perfect energy transfer takes place between the two modes $x=1$ and $x=2$ according to a timescale specified by the system Bohr frequency $2J$. Initially localized on the mode $x=1$, the vibrons reach the mode $x=2$ at the transfer time $\tau=\pi/2J$ and they recur on the excited site at time $2\tau$. 
%
% When $A=4$, a different behavior arises (see Fig.~\ref{fig:DP}). Although the energy does not localize, it takes an extremely long time to propagate from one mode to the other. 
%The dynamics is governed by different timescales. Over short time scales, $\Delta P(t)$ remains close to its initial value by exhibiting  high-frequencies oscillations around a mean value approximately equal to $2.9$. This feature originates from the non-resonant coupling between the initial state $|0,3,0\rangle$, with self-energy $-6A$, and the state $|0,2,1\rangle$ with self-energy $-2A$. %The gap between these states being equal to $4A$, it gives rise to oscillations with a period equal to $2\pi/4A=0.39$. 
%

For a non-vanishing vibron-photon coupling ($G > 0$), it turns out that a cavity-enhanced QST process takes place for small $G$ values.
Indeed, when $G = 1$ and $2$, the vibron population imbalance behaves as it did before when $G=0$, but now with increasing periods of oscillations. 
Notably, the transfer time increases to approximately $90$ for $G=1$ and it becomes longer than $250$ for $G=2$. 
Then, for $G=3$ the behavior of $\Delta P(t)$ changes completely since energy localization arises over the considered timescale resulting in a strongly reinforced QST.
In this case, $\Delta P(t)$ exhibits high-frequency oscillations around a mean value approximately equal to $2.62$ indicating that over $87\%$ of the vibrational energy remains localized where it was initially deposited. 
However, a fully different behavior is observed for larger $G$ values. Indeed, for $G=4$, a shorter transfer time recurs of approximately $125$. This effect is enhanced as $G$ increases so that a fast energy transfer takes place for $G=5$. The transfer time reaches approximately $15$. In that case, cavity-assisted energy transfer takes place resulting in a transfer time approximately of the order of magnitude of the transfer time for free vibrons. Nevertheless, the efficiency of the transfer is not perfect. The maximum value of the vibron number on the mode $x=2$ is approximately equal to $2.7$ for $G=5$.

The previous results suggest that very stable QST arises in the region $G \sim 3$. 
% In fact, this result is merely the product of a fortunate coincidence and we have verified that, over a longer time scale, an exchange occurs between the initial state and the target state, resulting in a transfer of vibrational energy between the two modes. 
% The transfer time is extremely long, reaching approximately $\tau=23000J^{-1}$. However,
By analyzing the dynamics near this point, we have identified the existence of a critical value for the light-matter coupling strength $G_c =3.01675$ for which a fully stabilized light-induced QST occurs (with parameters $N_B = 3$ and $A = 4$). 
As illustrated in Fig.~\ref{fig:COURBEGC} for $G=G_c$, the vibron population imbalance $\Delta P(t)$ initially equal to $3$ slightly decreases to exhibit high-frequency small-amplitude oscillations around a mean value approximately equal to~$2.7$.
Such a behavior is shown here over a timescale of $5\times10^4$, but we have numerically verified that it persists over an even longer timescale, up to $10^{15} $. 
This behavior is a direct signature of the arising of a stabilized light-induced~QST, in which approximately $90\%$ of the vibrational energy remains localized where it has been deposited. 
% As shown in Fig.~\ref{fig:COURBEGC}b, due to this genuine self-trapping, the vibrons are no longer able to reach the target state, the corresponding quantum probability $\pi_T(t)$ fluctuating around a very small value approximately equal to $1.6\times 10^{-3}$. Instead, the system remains confined in a region of the Hilbert space, mainly formed by the initial state, the corresponding survival probability oscillating around $0.82$. Nevertheless, the system is able to explore part of the Hilbert space corresponding to photon exchange. Indeed, the probabilities $\pi_2(t)$  and $\pi_0(t)$ show high-frequency modulations around mean values approximately equal to $0.15$ and $0.04$, respectively. 

Let us conclude here by mentioning an important point: the red curve in Fig.~\ref{fig:DP} and Fig.~\ref{fig:COURBEGC} defines the ``smoothed'' vibron population imbalance. It is numerically constructed by restricting the expansion of the evolution operator to the two lowest energy polaritonic eigenstates only (see Eq.~(\ref{eq:U})).
As readily seen, this curve perfectly captures the slowly varying behavior of the population imbalance, indicating that the two low-lying polaritonic eigenstates contain most of the relevant information to understand the energy transfer between the vibrational modes of the dimer.
This will be helpful when analyzing later on the signature of the light-induced QST in the polaritonic spectrum (see Sec.~\ref{sec:spectrum}).

\begin{figure}
%    \centering
    \includegraphics[width=8.5cm]{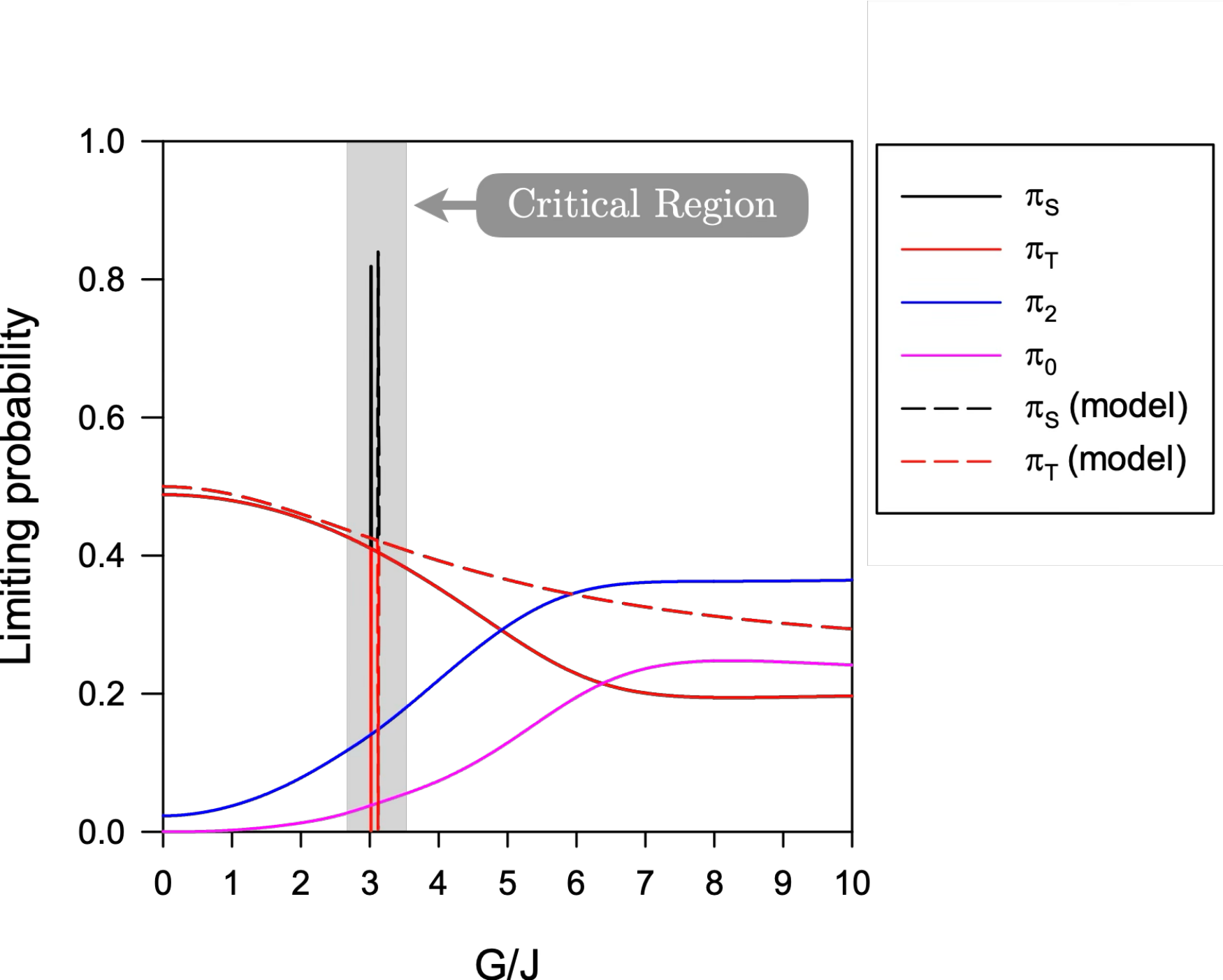}
    \caption{$G$ dependence of the limiting probabilities for $N_B=3$, $J=1$, $A=4$ and $\omega=\omega_0$. 
    The survival probability $\pi_S$ (black curve) is defined as the probability of finding the system in its initial state $|0,3,0\rangle$. The target probability $\pi_T$ (red curve) corresponds to the probability of finding the system in the target state $|0,0,3\rangle$. Finally, $\pi_2$ (blue curve) denotes the probability of occupying states with energy $3\omega_0 - 2A$, whereas $\pi_0$ (purple curve) denotes the probability of occupying states with energy $3\omega_0$ (see Fig.~\ref{fig:networks}).
    Full lines refer to numerical calculations whereas dashed lines correspond to values obtained with the four-level model defined in Sec.~\ref{sec:discussion}.}
\label{fig:LIMIT1}
\end{figure}

\begin{figure*}
    \centering
    \includegraphics[width=17cm]{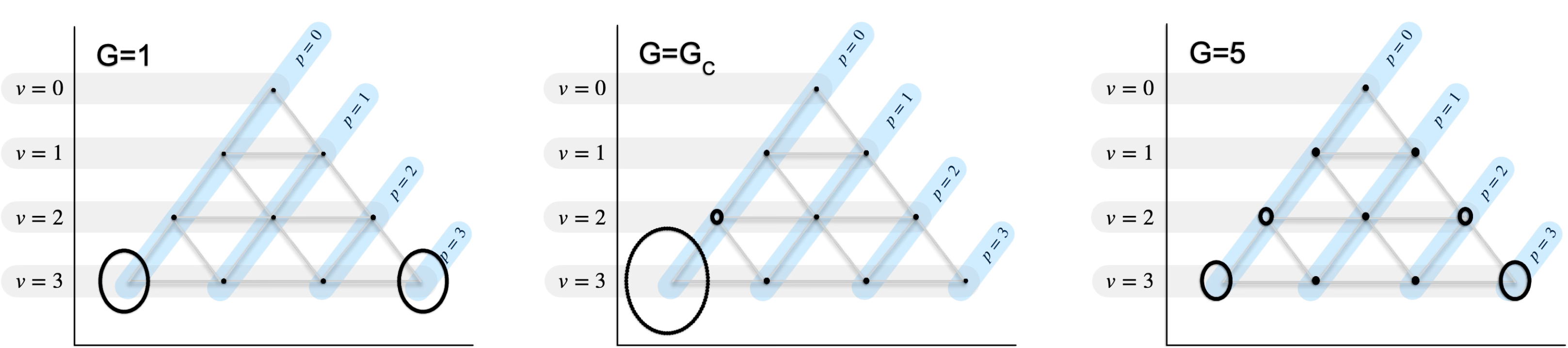}
    \caption{Two-dimensional representation of the limiting probabilities on the triangular network $\bar{\pi}_{v,p}^{(N_B)}$ for $N_B=3$, $J=1$, $A=4$, $\omega=\omega_0$, and for $G=1$, $G=G_c$, and $G=5$.}
\label{fig:LIMIT2}
\end{figure*}

\subsubsection{Limit Probabilities: \\  Light-Induced Symmetry Breaking }

We now show that the existence of a critical light–matter coupling at which QST becomes stabilized is linked to the emergence of symmetry breaking in the system. To highlight this connection, we consider the triangular lattice (see Fig.~\ref{fig:networks}) and evaluate the limiting probabilities, as defined in Eq.~(\ref{eq:limit_prob}), across the network, examining how they vary with the light–matter coupling..

% adopt a different perspective focusing
% Within this framework, the many-vibron transfer dynamics---consisting of the complete transfer of the $N_B$ vibrons from the left anharmonic mode to the right one---is directly mapped onto an effective single-particle problem, i.e., a quantum walker that starts from the bottom-left site of the triangular network (corresponding to the configuration $|0,3,0\rangle$) and propagates entirely to the target bottom-right site $|0,0,3\rangle$ (see Fig.~\ref{fig:networks}).
% Adopting this viewpoint allows us to probe which light--matter configurations, i.e., effective sites on the triangular network, play a key role in realizing the transfer.
In Fig.~\ref{fig:LIMIT1}, we illustrate the evolution of four characteristic limiting probabilities as a function of the light–matter coupling $G$. The survival probability $\pi_S$ (black curve) measures the probability that the system remains in its initial state $|0,3,0\rangle$, while the target probability $\pi_T$ (red curve) represents the probability of reaching the mirror state $|0,0,3\rangle$. The probabilities $\pi_2$ (blue curve) and $\pi_0$ (purple curve) correspond to occupation of states with energies $3\omega_0 - 2A$ and $3\omega_0$, respectively (see Fig.~\ref{fig:networks}). The sum of all four probabilities is always unity.
Away from the critical region (gray area), $\pi_S$ and $\pi_T$ are equal, reflecting the delocalized nature of the polaritonic eigenstates and the mirror symmetry of the AQD, which prevents stable self-trapping. At $G=0$, both probabilities start at 0.49 and decrease with increasing $G$, reaching roughly 0.2 in the strong coupling limit. Meanwhile, $\pi_2$ and $\pi_0$ grow with $G$, reaching 0.36 and 0.24, respectively, from initial values of 0.023 and 0.
At the critical coupling $G_c$ (see gray area), a sudden discontinuity appears for two curves: $\pi_S$ jumps to 0.82, while $\pi_T$ drops to $1.68\times 10^{-3}$. 
The other probabilities, $\pi_2=0.140$ and $\pi_0=0.038$, remain largely unaffected. 
This sharp change signals a symmetry-breaking transition, where the system abruptly favors the initial state $|0,3,0\rangle$ over its mirror state $|0,0,3\rangle$, fundamentally altering how it occupies the triangular network.

To complement these observations, in Fig.~\ref{fig:LIMIT2} we illustrate the symmetry-breaking transition  through a two-dimensional representation of the global limiting probabilities $\bar{\pi}_{v,p}^{(N_B)}$  on the triangular lattice for $N_B=3$, $A=4$, and couplings $G=1$, $G=G_c$, and $G=5$. 
Each site $(v,p)$ of the Hilbert-space network (see Fig.~\ref{fig:networks}) is marked by a circle whose radius corresponds to the probability of the configuration $|N_B-v, v-p, p\rangle$.

In the weak-coupling regime ($G=1$), the probabilities exhibit mirror symmetry, $\bar{\pi}_{v,p}^{(N_B)} = \bar{\pi}_{v,v-p}^{(N_B)}$, and are mainly concentrated on the initial $(v=3,p=0)$ and target $(v=3,p=3)$ states, both with values of 0.48. Other states, such as $(v=3,p=1)$, $(v=3,p=2)$, $(v=2,p=0)$, and $(v=2,p=2)$, have small probabilities around $9\times10^{-3}$. This pattern reflects oscillations between initial and target states mediated by higher-energy states, preventing stabilized QST. The exchange of a single photon slightly modifies the dynamics compared with $G=0$, increasing the transfer time.

At the critical coupling $G=G_c$, mirror symmetry is broken. The system localizes strongly on the initial state $(v=3,p=0)$, which reaches 0.82, while one-photon states such as $(v=2,p=0)$ rise modestly to 0.093. Other relevant states, including $(v=3,p=1)$ and $(v=3,p=2)$, remain around 0.015. Crucially, the target state probability drops to $1.6\times10^{-3}$. This behavior confirms the onset of fully stabilized light-induced QST, with all three vibrons confined to their initial site, marking a clear symmetry-breaking transition.

In the strong-coupling regime ($G=5$), the probabilities delocalize along the edges of the triangular network, $(v,p=0)$ and $(v,p=v)$ for $v=0,1,2,3$. The initial and target states remain dominant at 0.19, while one- and two-photon states, such as $(v=2,p=0)$ and $(v=1,p=0)$, increase to 0.12 and 0.07, respectively. Even the vibron-free state $(v=0,p=0)$ gains a small probability of 0.03. This shows broader participation of all light-matter configurations across the network while maintaining mirror symmetry.

\begin{figure*}
    \centering
    \includegraphics[width=17cm]{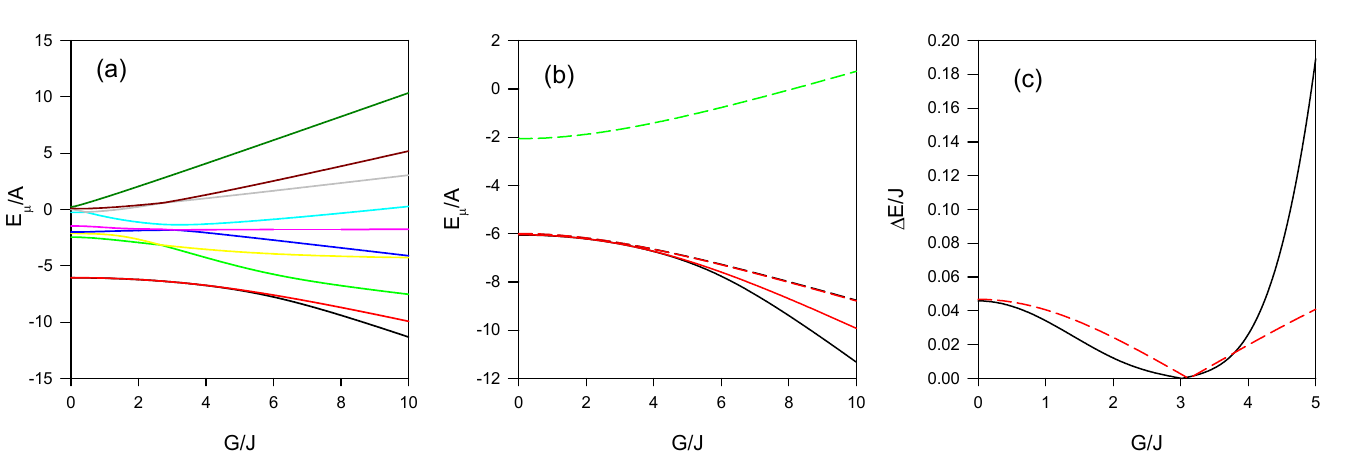}
    \caption{(a) $G$ dependence of the polaritonic energy spectrum. (b) Dashed lines correspond to $G$ dependence of the energy spectrum of the four level model (see Sec.~\ref{sec:discussion}). Black and red full lines define the two lowest energy level of the polaritonic energy spectrum extracted form numerical calculations. (c) $G$ dependence of the gap between the two lowest energy level form numerical calculation (black full line) and from the four-lvel model (red dahsed line). The parameters are $N_B=3$, $J=1$, $A=4$, and $\omega=\omega_0$.}
\label{fig:SPECTRA}
\end{figure*}

\subsubsection{Polaritonic Energy Spectrum: \\ Spectral Signature of Light-Induced QST}
\label{sec:spectrum}

To understand the origin of the critical point, let us investigate the nature of the polaritonic energy spectrum whose dependence on $G$ is shown in Fig.~\ref{fig:SPECTRA}a. According to the shape of the Hamiltonian displayed in Fig.~\ref{fig:networks}, when $G=0$, the full energy spectrum includes the spectra of the AQD when 0, 1, 2, or 3 vibrons are present. Due to the degeneracy of the self-energies, the spectra overlap. For example, for $v=3$, the dimer eigenenergies occur around $3\omega_0-6A$, and $3\omega_0-2A$, whereas for $v=2$, they occur around $3\omega_0-2A$, and $3\omega_0$. Therefore, in the region around $3\omega_0-2A$, the full spectrum shows contributions from the dimer spectrum for $v=3$ and $v=2$, respectively. As $G$ increases, couplings occur between spectra with nearest-neighbor $v$ values, resulting in hybridization and avoid-crossing mechanisms. In accordance with the triangular lattice shown in Fig.~\ref{fig:networks}, these mechanisms first affect high energy states ($3\omega_0-2A$ and $3\omega_0$ are four time degenerated).  

To explore the QST process, special attention must be given to the lowest energy eigenstates which account for the coupling between the configurations $|0,3,0\rangle$ and $|0,0,3\rangle$ when $G=0$ (energy $\omega_0-6A$). Fig.~\ref{fig:SPECTRA}a suggests that these two states whose energy decreases as $G$ increases, are almost degenerated. The degeneracy seems to be lifted only for strong coupling. However, Fig.~\ref{fig:SPECTRA}c shows a more detailed analysis of the two lowest energy states, displaying the $G$ dependence of the gap $\Delta E = E_1 - E_0$ between them (black full line). One observes a quite surprising behavior since different regimes take place. 

Indeed, when $G=0$, standard perturbation theory shows that $\Delta E \approx 2|J_{\mathrm{eff}}(3)|$ (see Sec. II.C), where $J_{\mathrm{eff}}(3)$ is the effective coupling between the states $|0,3,0\rangle$ and $|0,0,3\rangle$. This effective coupling leads to oscillations of the vibrational energy between the two modes, the transfer time being approximately equal to $\tau = \pi/\Delta E$. As $G$ increases from zero, Fig.~\ref{fig:SPECTRA}c shows that the two lowest energy levels first move towards each other. The gap $\Delta E$ decreases as $G$ increases resulting in slowdown of the energy exchange between the two modes: the transfer time increases with $G$. This behavior continues until $\Delta E$ reaches a minimum value approximately equal to $10^{-15}$ for a critical coupling strength $G_c= 3.016749$. In other words, the gap between the two lowest energy states vanishes $\Delta E \rightarrow 0$ at the critical point so that the original degeneracy recurs. 
A fully stabilized QST occurs, characterized by an infinitely long transfer time. 
Note that around the critical point, the gap scales linearly with the coupling strength following the law $\Delta E \approx 8.3 \times 10^{-3} |G-G_c|$. Then, for larger values of $G$, the gap increases rapidly. This leads to an acceleration of the dynamics, which become faster and faster as $G$ increases, resulting in a decrease of the transfer time between the two modes.

\begin{figure}
%    \centering
    \includegraphics[width=7.5cm]{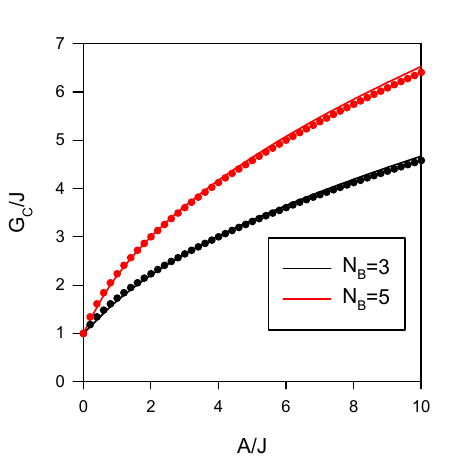}
    \caption{$A$ dependence of the critical coupling $G_c$ for $J=1$, $\omega=\omega_0$, $N_B=3$ (black curve) and $N_B=5$ (red curve). Cicrles characterize the empirical law $G_c=\sqrt{J^2+(N_B-1)AJ}$.}
\label{fig:GCVSA}
\end{figure}

The previous results show that when $3$ vibrons are initially created in the mode $x=1$, they preferentially remain localized where they were deposited when $G=G_c$, thus indicating the existence of a stable energy localization.
However, a fundamental question remains unanswered: is this effect a general property, or does it depend on the dimer anharmonicity $A$ (\textit{i.e.} internal many-body interactions) and the number of bosons $N_B$?

To address this question, we first study the influence of the anharmonicity through Fig.~\ref{fig:GCVSA}, which shows the $A$ dependence of the critical coupling for $N_B=3$ (black curve) and $N_B=5$ (red curve). Whatever $N_B$, $G_c=1$ in the harmonic limit ($A=0$).  Then, when the anharmonicity switches on, $G_c$ increases with $A$. The larger the boson number, the faster the critical coupling increases. For instance, when $A=2$, one obtains  $G_c\approx2.21$ for $N_B=3$ and  $G_c\approx3.00$ for $N_B=5$. Similarly, for a stronger anharmonicity equal to $A=6$, one obtains $G_c\approx3.65$ and $ 5.08$ for $N_B=3$ and $5$, respectively. As shown by the circles in Fig.~\ref{fig:GCVSA},  this behavior is relatively well captured by the empirical law $G_c=\sqrt{J^2+(N_B-1)AJ}$.

\begin{table}[t]
\centering
\begin{ruledtabular}
\begin{tabular}{ccc}
$N_B$ & $G_c/J$ & $\Delta E /J$ \\
\hline
\\
1 & 1.0     & $\sim 10^{-16}$ \\
2 & 2.18802 & $\sim 10^{-2}$ \\
3 & 3.01675 & $\sim 10^{-15}$ \\
4 & 3.61527 & $\sim 10^{-5}$\\
5 & 4.17196 & $\sim 10^{-15}$\\
6 & 4.62356 & $\sim 10^{-9}$\\
7 & 5.07023 & $\sim 10^{-15}$ \\
8 & 5.50999 & $\sim 10^{-14}$ \\
\end{tabular}
\end{ruledtabular}
\caption{$N_B$ depenence of the critical coupling $G_c$ for $J=1$, $A=4$ and $\omega=\omega_0$}
\label{tab:example}
\end{table}

Finally, the $N_B$ dependence of the critical coupling $G_c$ is reported in Table I. The table reveals a very surprising effect: the occurrence of the light-induced QST strongly depends on the parity of $N_B$. 
More precisely,
when the number of bosons is odd, numerical simulations show that there always exists a critical coupling at which the two lowest-energy levels cross. As $N_B$ increases, this critical value also increases. In contrast, the situation is more subtle when $N_B$ is even. In this case, we observed that the gap between the two lowest-energy levels reaches a minimum when $G$ reaches a critical value. However, this minimum is not zero, indicating that there is no explicit level crossing. For example when $N_B=2$, the gap between the two lowest levels at the critical point is equal to $\Delta E = 8.32\times 10^{-2}$. Nevertheless, the larger $N_B$ becomes, the smaller the gap at the critical point, eventually becoming vanishingly small when typically $N_B \geq 8$.

At this stage, let us mention that numerical simulations have been carried for an even number of bosons $N_B=2$ and $N_B=4$. In both cases one recovers the two asymptotic regimes, namely ;ight-enhanced quantum self-trapping in the weak coupling limit and cavity-assisted energy transfer in the strong coupling limit. Nevertheless, we did not observe the occurrence of a fully stabilized QST.
For even $N_B$ value, as listed in Table I, a critical coupling $G_c$ can be defined as the value of $G$ for which the transfer time is maximum. But it is not infinite as the transfer time obtained for odd $N_B$ values. The transfer time is approximately equal to $\tau=37.76$ for $N_B=2$ and it reaches $\tau=9.2\times10^{4}$ for $N_B=4$.  The key point is that, even though exact vibrational energy localization does not occur for even values of $N_B$, the interaction with the cavity still allows an optimization of the quantum self-trapping process compared to the case $G=0$. This effect becomes particularly significant for large boson numbers, so that for $N_B>8$ one can effectively speak of a fully stabilized light-induced QST.

\section{Discussion and Interpretation}
\label{sec:discussion}

The numerical results reveal that the hybridization between the electromagnetic mode and the vibrational modes gives rise to specific dynamical behaviors. Starting from an initial state involving $N_B$ vibrons localized 
on mode $x = 1$, we observe that the dynamics is mainly governed by the two lowest polaritonic energy levels. The energy gap between these levels measures the effective hopping constant  that controls the transfer between the initial state and the target state involving $N_B$ vibrons on mode $x = 2$. In that context, the analysis of the quantum probabilities shows that increasing the vibron-photon coupling strength $G$ enlarges the number of pathways in Hilbert space connecting  initial and target states. Quantum interferences between these pathways then modify the effective coupling between these states and two different regimes take place. 
In the weak-coupling regime, the energy gap between the two-lowest energy levels decreases with $G$, leading to a slowdown of 
the vibrational energy transfer,  a phenomenon referred to as photon-enhanced quantum self-trapping. In contrast, in the strong-coupling regime, 
the opposite behavior occurs: the energy gap increases with $G$, resulting in a very fast energy transfer, i.e. cavity-assisted energy transfer. A key result is the
existence of a critical region in parameter space where the  two lowest-energy levels intersect without displaying an avoided crossing. In this situation, the effective hopping constant between the initial and target states vanishes  and the vibrational energy becomes localized, remaining predominantly trapped on the mode where it was created. This corresponds to the onset of the stabilized QST. 

%Note that degeneracy-induced localization is a well-established mechanism in complex networks, where excitons may localize depending on the network symmetry~\cite{mulken06a}. In the present situation, it results from the isomorphism between the AQD dynamics and that of a fictitious particle moving on a triangular network.

\subsection{The Four-level model} 

To interpret the previous numerical observations, let us introduce a simple phenomenological model. To proceed, the main idea is to restrict our analysis to an active subspace containing the states that are relevant to the dynamics, while neglecting the influence of the remainig irrelevant states. These relevant states depend on the initial conditions chosen to describe the dynamics. In the present work, we consider the initial creation of $N_B$ vibrons on mode $x=1$, with the cavity initially in the photon-free state and mode $x=2$ occupying its zero vibron ground state. In this situation, when the coupling to the cavity is switched off, the states relevant to QST are $\ket{0,N_B,0}$ and $\ket{0,0,N_B}$. As mentioned in Sec.~\ref{sec:QD}, these two states are indirectly coupled through their interaction with the remaining states of the $N_B$-vibron subspace, resulting in an effective hopping constant $J_{\mathrm{eff}}(N_B)$ (see Eq.(\ref{eq:Jeff}))). In other words, the active substance, disregarding the vibron--photon interaction, corresponds to an effective two-level system involving the two lowest energy levels of the AQD.

When $G\neq0$, the two states $\ket{0,N_B,0}$ and $\ket{0,0,N_B}$ become coupled to the states $\ket{1,N_B-1,0}$ and $\ket{1,0,N_B-1}$, through the creation or annihilation of a photon in the cavity mode. In the $(N_B-1)$-vibron subspace, the two states $\ket{1,N_B-1,0}$ and $\ket{1,0,N_B-1}$ define the active two-level system involved in the self-trapping process in the presence of $(N_B-1)$ vibrons. These two states are indirectly coupled through their interaction with the remaining states of the $(N_B-1)$-vibron subspace, giving rise to an effective hopping constant $J_{\mathrm{eff}}(N_B-1)$.

\begin{figure}[t]
    \centering
    \includegraphics[width=\columnwidth]{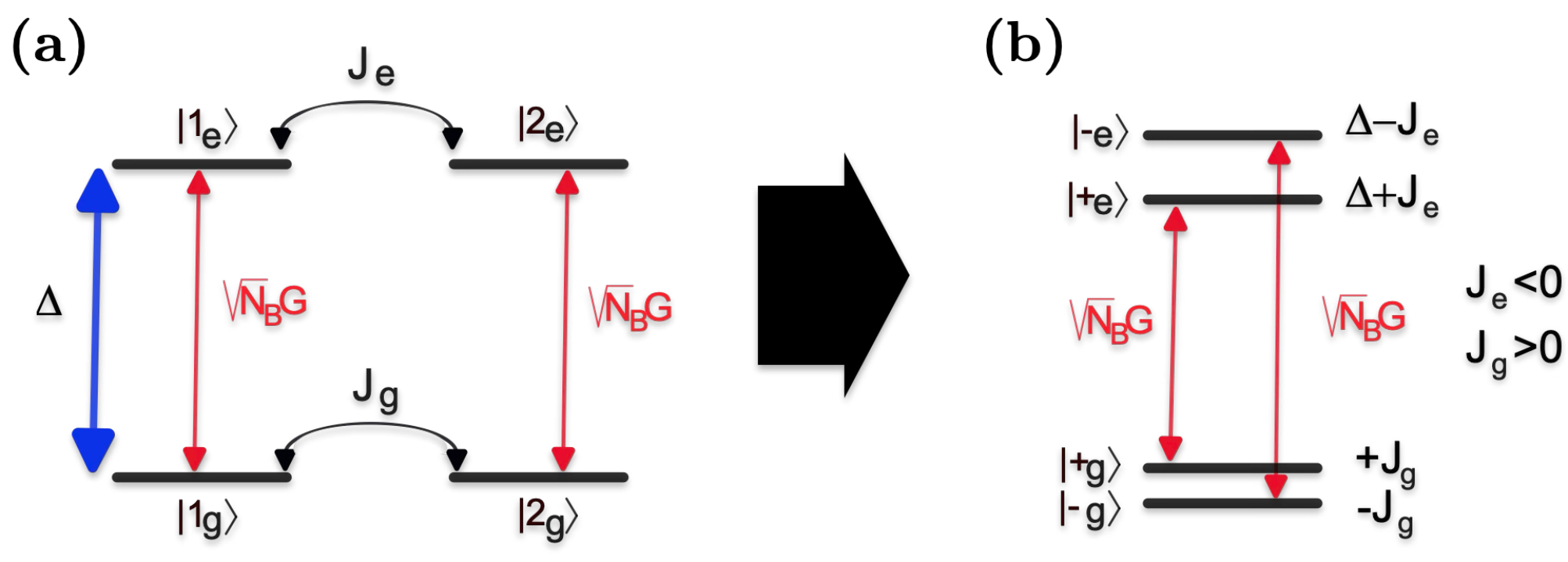}
    \caption{(a) Four-level model introduced in Sec.~\ref{sec:discussion} to capture the physics of the NQD in the cavity. It represents the active subspace formed by combining the two-level system spanned by the states $\ket{1_g}=\ket{0,N_B,0}$ and $\ket{2_g}=\ket{0,0,N_B}$ and the two-level system spanned by the states $\ket{1_e}=\ket{1,N_B-1,0}$ and $\ket{2_e}=\ket{1,0,N_B-1}$.  
(b) Illustration of the couplings between the symmetric states and between the antisymmetric states (see Sec.~\ref{sec:discussion}). }
    \label{fig:four_level}
\end{figure}

Consequently, as illustrated in Fig.~\ref{fig:four_level}a, the active subspace reduces to a four-level system formed by combining, on the one hand, the two-level system spanned by the states $\ket{1_g}=\ket{0,N_B,0}$ and $\ket{2_g}=\ket{0,0,N_B}$, and, on the other hand, the two-level system spanned by the states $\ket{1_e}=\ket{1,N_B-1,0}$ and $\ket{2_e}=\ket{1,0,N_B-1}$. Owing to the coupling with the cavity, $\ket{1_g}$ (respectively $\ket{2_g}$) interacts with $\ket{1_e}$ (respectively $\ket{2_e}$), with an interaction strength $\sqrt{N_B}\,G$. Note that we will denote by $J_g = J_{\mathrm{eff}}(N_B)$ (respectively, $J_e = J_{\mathrm{eff}}(N_B-1)$) the effective hopping constant in the low-energy two-level system (respectively, in the high-energy two-level system). As shown in Fig.~\ref{fig:four_level}a, we have introduced the parameter $\Delta= 2 (N_B-1) A$, which denotes the energy difference
$E^{(N_B-1)}_0-E^{(N_B)}_0$ between the two-level system defined in the $(N_B-1)$-vibron subspace and that defined in the $N_B$-vibron subspace (see Eq.(\ref{eq:EJG})).

In the basis $\{ \ket{1_g}, \ket{2_g}, \ket{1_e}, \ket{2_e} \}$, the Hamiltonian of the four-level system is defined as
\begin{equation}
\mathbb{H} =
\begin{pmatrix}
0     & J_g   & \sqrt{N_B}\,G & 0 \\
J_g   & 0     & 0             & \sqrt{N_B}\,G \\
\sqrt{N_B}\,G & 0 & \Delta & J_e \\
0 & \sqrt{N_B}\,G & J_e & \Delta
\end{pmatrix}.
\end{equation}
To diagonalized $\mathbb{H}$, we first introduce the basis $\ket{\pm_g}$, which yields a diagonal representation of the two-level system Hamiltonian defined in the $N_B$-vibron subspace. The corresponding eigenvectors, associated with the eigen-energies $\pm J_g$, are defined as
\begin{equation}
\ket{\pm_g} = \frac{1}{\sqrt{2}} \left( \ket{1_g} \pm \ket{2_g} \right).
\end{equation}
We then define the basis $\ket{\pm_e}$, which diagonalizes the two-level system Hamiltonian defined in the $(N_B-1)$-vibron subspace. Associated to the eigen-energies $\Delta \pm J_e$, these eigenvectors read
\begin{equation}
\ket{\pm_e} = \frac{1}{\sqrt{2}} \left( \ket{1_e} \pm \ket{2_e} \right).
\end{equation}
Due to both the symmetry of the eigenstates and the nature of the vibron--photon interaction, the coupling with the cavity mode induces an interaction between $\ket{+_g}$ and $\ket{+_e}$ on the one hand, and between $\ket{-_g}$ and $\ket{-_e}$ on the other hand. As a result, the four-level Hamiltonian becomes block diagonal and can be rewritten as (see Fig.~\ref{fig:four_level}b)
\begin{equation}
\mathbb{H} =
\begin{pmatrix}
\mathbb{H}_s & 0 \\
0   & \mathbb{H}_a
\end{pmatrix}.
\end{equation}
The block $\mathbb{H}_s$ is the restriction of the Hamiltonian to the symmetric subspace spanned by the vectors $\ket{+_g}$ and $\ket{+_e}$.
It is defined as
\begin{equation}
\mathbb{H}_s =
\begin{pmatrix}
J_g & \sqrt{N_B}\,G \\
\sqrt{N_B}\,G & \Delta + J_e
\end{pmatrix}.
\end{equation}
Similarly, the block $\mathbb{H}_a$ is the restriction of the Hamiltonian to the antisymmetric subspace spanned by the vectors $\ket{-_g}$ and $\ket{-_e}$.
It is defined as
\begin{equation}
\mathbb{H}_a =
\begin{pmatrix}
- J_g & \sqrt{N_B}\,G \\
\sqrt{N_B}\,G & \Delta - J_e
\end{pmatrix}.
\end{equation}
%
%
%At this stage, without loss of generality, we assume $J_e < 0$ and $J_g > 0$.
%This situation occurs when $N_B$ is odd.
%The case of even values of $N_B$ can be recovered straightforwardly.
%Moreover, to simplify the notations, we introduce the following parameters:
%\begin{equation}
%\varepsilon = \frac{\Delta}{2}, \qquad 
%\bar{J} = \frac{|J_e| + J_g}{2}, \qquad
%\delta J = \frac{|J_e| - J_g}{2}.
%\end{equation}
%
%With these definitions, the eigenvalues of $\mathbb{H}_s$ and $\mathbb{H}_a$, called $\omega_s^{(\pm)}$ and $\omega_a^{(\pm)}$ respectively, are defined as 
%\begin{align}
%\omega_s^{(\pm)}
%&= \varepsilon - \delta J
%\pm \sqrt{(\varepsilon - \bar{J})^2 + N_B G^2}, \\ \nonumber
%\omega_a^{(\pm)}
%&= \varepsilon + \delta J
%\pm \sqrt{(\varepsilon + \bar{J})^2 + N_B G^2}.
%\end{align}
 
The energy spectrum of the four-level system is determined by the eigenvalues of $\mathbb{H}_s$ and $\mathbb{H}_a$, denoted $\omega_s^{(\pm)}$ and $\omega_a^{(\pm)}$, respectively. They read
\begin{eqnarray}
\omega_s^{(\pm)} &=& \frac{\Delta+J_e+J_g}{2} \pm \sqrt{\left( \frac{\Delta+J_e-J_g}{2} \right)^2 + N_B G^2} \nonumber \\
\omega_a^{(\pm)} &=& \frac{\Delta-J_e-J_g}{2} \pm \sqrt{\left( \frac{\Delta-J_e+J_g}{2} \right)^2 + N_B G^2}  \nonumber \\
\end{eqnarray}
The dependence of these four eigenenergies on $G$ is shown in Fig.~\ref{fig:SPECTRA}b (dashed lines). Their relative ordering depends on the parity of the vibron number $N_B$.

For odd $N_B$ ($J_g>0$ and $J_e<0$), and at $G=0$, the levels are ordered as
$\omega_a^{(-)}<\omega_s^{(-)}<\omega_s^{(+)}<\omega_a^{(+)}$.
As $G$ increases, avoided crossings occur within each symmetry sector, namely between $\omega_s^{(-)}$ and $\omega_s^{(+)}$, and between $\omega_a^{(-)}$ and $\omega_a^{(+)}$. Consequently, $\omega_s^{(-)}$ and $\omega_a^{(-)}$ decrease with increasing $G$, while $\omega_s^{(+)}$ and $\omega_a^{(+)}$ increase.
However, since $J_g>0$ and $J_e<0$ (see Fig.~\ref{fig:four_level}b), the energy gap $\Delta+J_e-J_g$ between $\ket{+_g}$ and $\ket{+_e}$ is smaller than the gap $\Delta-J_e+J_g$ between $\ket{-_g}$ and $\ket{-_e}$. As a result, the light--matter interaction affects the symmetric sector more strongly than the antisymmetric one. In particular, $\omega_s^{(-)}$ decreases faster with $G$ than $\omega_a^{(-)}$, leading to a critical value of $G$ at which the two levels cross without hybridizing.
For even $N_B$ ($J_g<0$ and $J_e>0$), the situation is reversed: the antisymmetric sector is more strongly affected, and $\omega_a^{(-)}$ decreases faster than $\omega_s^{(-)}$. The critical point still corresponds to the crossing of the two lowest levels.

Therefore, irrespective of the parity of $N_B$, the critical coupling $G_c$ is obtained from the condition $\omega_s^{(-)} = \omega_a^{(-)}$, yielding
\begin{equation}\label{eq:Gcritic}
N_B G_c^2 = -\frac{J_g J_e}{(J_g+J_e)^2} \left( \Delta^2 - (J_g+J_e)^2 \right).
\end{equation}
For $N_B=3$, Eq.~(\ref{eq:Gcritic}) gives $G_c \approx 3.12$, in good agreement with the exact value $G_c = 3.01675$. As noted above, the existence of the degeneracy requires $J_g J_e < 0$. Moreover, in the limit $A \gg J$, Eq.~(\ref{eq:Gcritic}) shows that $G_c$ scales as $\sqrt{J A}$, in agreement with Fig.~\ref{fig:GCVSA}. However, this expression does not capture how $G_c$ depends on the parameter $N_B$.

\subsection{Quantum Dynamics}

The evolution operator $\mathbb{U}(t) = \exp(-i \mathbb{H} t)$ which governs the four-level dynamics is defined in terms of a contour integral on a closed path $\Gamma$ in the complex plane as~\cite{cohen98}
\begin{equation}\label{eq:UG}
    \mathbb{U}(t) = \frac{1}{2 i \pi} \oint_\Gamma \mathbb{G}(z) e^{-i z t} \, dz,
\end{equation}
where $\mathbb{G}(z)=(z-\mathbb{H})^{-1}$ is the Green function. The closed path $\Gamma$ is formed by a straight line located just above the real axis and closed by a half-circle turned towards the negative imaginary axis. The radius of the half-circle tends to infinity to apply the residue theorem. The Green function has the same block-diagonal structure as the four-level Hamiltonian:
\begin{equation}
    \mathbb{G} = 
    \begin{pmatrix}
        \mathbb{G}_s & 0 \\
        0 & \mathbb{G}_a
    \end{pmatrix},
\end{equation}
where $\mathbb{G}_{s,a}(z)=(z-\mathbb{H}_{s,a})^{-1}$ is the restriction of the whole Green function to the symmetric subspace and to the antisymmetric subspace. From the expression of $\mathbb{H}_s$ in the basis $\{|+_g\rangle, |+_e\rangle\}$, $\mathbb{G}_s$ reads
\begin{equation}
    \mathbb{G}_s(z) = \frac{1}{\Delta_s(z)}
    \begin{pmatrix}
        z - \Delta - J_e & \sqrt{N_B} G \\
        \sqrt{N_B} G & z - J_g
    \end{pmatrix}.
\end{equation}
where $\Delta_s(z)=(z-J_g)(z-\Delta-J_e) - N_B G^2$. Similarly, from the expression of $\mathbb{H}_a$ in the basis $\{|-_g\rangle, |-_e\rangle\}$, $\mathbb{G}_a$ reads
\begin{equation}
    \mathbb{G}_a(z) = \frac{1}{\Delta_a(z)}
    \begin{pmatrix}
        z - \Delta + J_e & \sqrt{N_B} G \\
        \sqrt{N_B} G & z + J_g
    \end{pmatrix}.
\end{equation}
where $\Delta_a(z)=(z+J_g)(z-\Delta+J_e) - N_B G^2$.

To explain the occurrence of a fully stabilized QST, let us consider that the system is initially in the state $|1_g\rangle = |0, N_B, 0\rangle$. The survival amplitude $\mathbb{U}_S(t)$, i.e. the probability amplitude of observing the system in its initial state at time $t$, is defined as $\mathbb{U}_S(t)=\langle 1_g | \mathbb{U}(t) | 1_g \rangle$. The target amplitude $\mathbb{U}_T(t)$, i.e. the probability amplitude of observing the system in the target state $|2_g\rangle = |0,0,N_B\rangle$ at time $t$, is defined as $\mathbb{U}_T(t)=\langle 2_g | \mathbb{U}(t) | 1_g \rangle$. From the knowledge of the Green function, Eq.(\ref{eq:UG}) can be solved and one obtains
\begin{align}
\mathbb{U}_S(t) &= \alpha_+ e^{-i \omega_s(+) t}
       + \alpha_- e^{-i \omega_s(-) t} \notag \\
       &+ \beta_+  e^{-i \omega_a(+) t}
       + \beta_-  e^{-i \omega_a(-) t}, \notag \\
\mathbb{U}_T(t) &= \alpha_+ e^{-i \omega_s(+) t}
       + \alpha_- e^{-i \omega_s(-) t} \notag \\
       &- \beta_+  e^{-i \omega_a(+) t}
       - \beta_-  e^{-i \omega_a(-) t}.
\end{align}
where  the various coefficients are defined as
\begin{align}
\alpha_\pm &= \frac{1}{4} \left( 1 \mp \frac{\Delta+J_e-J_g}{\sqrt{(\Delta+J_e-J_g)^2 + 4N_B G^2}} \right), \notag \\
\beta_\pm  &= \frac{1}{4} \left( 1 \mp \frac{\Delta-J_e+J_g}{\sqrt{( \Delta-J_e+J_g)^2 + 4N_B G^2}} \right).
\end{align}

\begin{figure*}
    \centering
    \includegraphics[width=18cm]{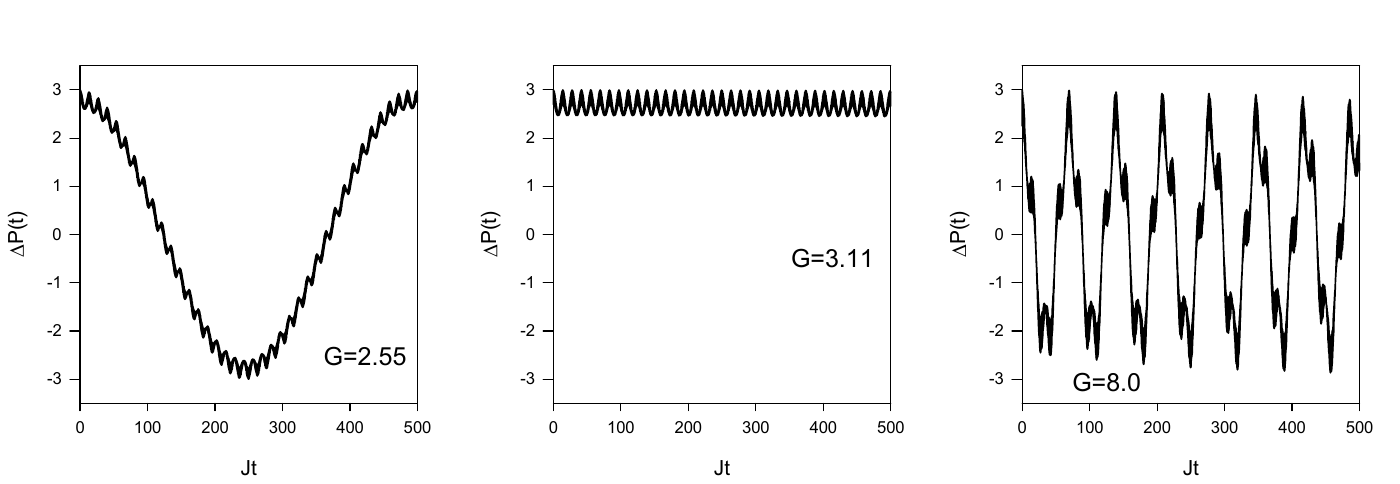}
    \caption{Time evolution of the vibron population imbalance $\Delta P(t)$ from the four-level model for $N_B=3$, $J=1$ $A=4$ and $\omega=\omega_0$. Three $G$ values are considered to illustrate the different dynamical regimes.}
    \label{fig:pi_theory}
\end{figure*}
 
Based on these quantities, we can evaluate the time evolution of the vibron population imbalance $\Delta P(t)$, as shown in Fig.~\ref{fig:pi_theory} for different values of the light–matter coupling.
The results, in qualitative agreement with the exact solution, reveal that increasing $G$ initially slows the energy exchange between the two vibrational modes, thereby enhancing QST. For larger $G$, however, this trend reverses, leading instead to cavity-assisted energy transfer. Notably, the figure highlights the arising of a stabilized light-induced QST at the critical coupling $G_c$.

From the analytical expressions of the probability amplitudes, one can extract the corresponding limiting probabilities. Two situations arise depending on the value of the light--matter coupling. Outside the critical region, the survival and target limiting probabilities are identical and read
\begin{equation}
\pi_S = \pi_T
= |\alpha_+|^2 + |\alpha_-|^2 + |\beta_+|^2 + |\beta_-|^2.
\end{equation}
By contrast, at the critical point where $\omega_s(-) = \omega_a(-)$, the survival and target limiting probabilities behave differently:
\begin{align}
\pi_S &= |\alpha_+|^2 + |\beta_+|^2 + |\alpha_- + \beta_-|^2, \notag \\
\pi_T &= |\alpha_+|^2 + |\beta_+|^2 + |\alpha_- - \beta_-|^2.
\end{align}
As shown in Fig.~\ref{fig:LIMIT1} (dashed lines), the four-level model provides quite good estimates of the $G$ dependence of the limiting probabilities. It captures the qualitative behavior of the probabilities, showing that both $\pi_S$ and $\pi_T$ are continuous decreasing functions of the coupling $G$ (equal to $0.5$ for $G=0$), except at the critical point where a discontinuity occurs. At this point, the limiting survival probability suddenly increases to reach $0.84$, whereas the limiting target probability drops to $3.68 \times 10^{-3}$, in very good agreement with the numerical results. 

\subsection{Interpretation}

The previous observations can be interpreted as follows. In the limit $G = 0$, to join the initial state $|1_g\rangle=|0,N_B,0\rangle$  and the target state $|2_g\rangle=|0,0,N_B\rangle$, the system must follow a path involving a series of states in the $N_B$-vibron subspace: $|0,N_B - 1,1\rangle$  , $|0,N_B - 2,2\rangle$, …, $|0,1,N_B-1\rangle$. This path is characterized by a transition amplitude whose strength is proportional to the effective hopping constant, $J_g=J_{\mathrm{eff}}(N_B)$. A measure of the transition amplitude is thus given by the gap between the two lowest energy levels $\Delta E = 2|J_g|$.

As $G$ switches on, new paths are created to connect the two relevant states, paths for which the number of photons in the cavity increases from zero. The probability amplitude to join the relevant states is the sum of the probability amplitudes associated with the various paths. Consequently, the gap between the two lowest energy levels exhibits a series of terms, each measuring the probability amplitude associated with a particular path. In the weak coupling limit, the gap $\Delta E$ behaves as
\begin{equation}
\Delta E \simeq 2 |J_g|
- \frac{N_B |J_e - J_g|}{2 (N_B - 1)^2 A^2} \, G^2 .
\label{eq:DELTAE_th}
\end{equation}
According to Eq.(\ref{eq:DELTAE_th}), $\Delta E$ decreases with $G$, in agreement with the numerical simulations. This leads to a slowdown of the energy transfer between the two vibrational modes, i.e. to an enhancement of the quantum self-trapping, which results from the quantum interferences between the different paths followed by the system to jump from the initial state to the target state. The dynamics slowdown originates from the difference between the two effective hopping constants $J_g$ and $J_e$. Note that for $G=1$, $N_B=3$, and $A=4$, Eq.(\ref{eq:DELTAE_th}) yields a transfer time  $\tau = 78$, in relatively good agreement with the numerical observation  ($\tau \approx 90$).

As  $G$ increases, the expansion of $\Delta E$ with respect to $G$ exhibits an alternating series of positive and negative contributions. This feature accounts for quantum interference between the different paths followed by the system. Depending on whether the interference is constructive or destructive, the full effective hopping constant first decreases as $G$ increases from zero, and then increases with strong $G$ values. A critical point exists in the parameter space for which $\Delta E$ vanishes: quantum interferences almost stops the dynamics resulting in the localization of the main part of the vibrational energy where it has been deposited: a stable QST takes place. Note that degeneracy-induced localization is a well-established mechanism in complex networks, where excitons may localize depending on the network symmetry~\cite{mulken06a}. In the present situation, the stable light-induced QST can be seen as the consequence from the isomorphism between the AQD dynamics and that of a fictitious single-particle moving on a triangular network.

In the strong coupling limit, the four-level model is 
not accurate enough to reproduce the dynamics in this regime because it is 
restricted to the exchange of only one photon, whereas one expects that 
several photons may participate. Nevertheless, it allows us to understand the 
physics involved in the process. Indeed, in the strong coupling limit, the two 
lowest energy levels become
\begin{align}
\omega_s^{(-)} &\approx\; \frac{\Delta+J_e+J_g}{2} - \sqrt{N_B}\,G, \notag \\
\omega_a^{(-)} & \approx\; \frac{\Delta-J_e-J_g}{2} + \delta J - \sqrt{N_B}\,G.
\end{align}
The corresponding energy gap is therefore $\Delta E \approx |J_e|$. This expression shows that the transfer time essentially depends 
on the effective coupling between the bound states involving $(N_B - 1)$  vibrons. 
The energy transfer is therefore assisted by the presence of a photon.  
Indeed, when $G$ becomes sufficiently strong, the initial state 
$|1_g\rangle $ couples to the state $|1_e\rangle $. This coupling favors the formation of two polaritonic states $|{po}_1^{(\pm)}\rangle = (|1_g\rangle \pm |1_e\rangle)/\sqrt{2}$,
 involving the vibrons of mode 1. The same phenomenon occurs on mode 2, resulting in the formation of two polaritonic states  $|{po}_2^{(\pm)}\rangle = (|2_g\rangle \pm |2_e\rangle)/\sqrt{2}$. The vibrational energy transfer between the two vibrational modes thus results from the coupling between the polaritonic states $|{po}_1^{(\pm)}\rangle$ and $|{po}_2^{(\pm)}\rangle$. In contrast to the transfer in standard quantum self-trapping, which depends only on the effective  coupling $J_g$, the interaction between the polaritonic states involves both effective hopping constants $J_g$ and $J_e$. Consequently, the acceleration of vibrational energy transfer in the strong coupling regime results from the fact that $|J_e| \gg |J_g|$.

To conclude, the four-level model provides a qualitative understanding of the dynamics of the AQD embedded in an optical cavity. It allows us to interpret most of our numerical observations and provides a physical picture of the different dynamical regimes. However, it remains too simple to capture all the underlying physics and is not quantitatively accurate. The discrepancy between the model and the exact calculations arises from the neglect of resonances between certain states, which are expected to significantly affect the system dynamics. For example, for $N_B=3$, resonances involve the states $\ket{1,0,2}$ and $\ket{1,2,0}$ included in the four-level model; these states are resonantly coupled to $\ket{0,2,1}$ and $\ket{0,1,2}$, respectively. Moreover, as pointed out earlier, the model cannot account for the strong coupling regime, since it only includes one-photon processes.

\section{Conclusion} 
\label{sec:conclu}

In this work, we have investigated how the energy exchange between two coupled anharmonic vibrational modes of a AQD is modified when the dimer is embedded in an optical cavity. The vibrational dynamics was described by an attractive Bose-Hubbard Hamiltonian, and the vibron-photon interaction was modeled using a single-mode Tavis-Cummings Hamiltonian. Special attention was devoted to characterizing the transition between an initial state consisting of $N_B$ vibrons localized on mode $x=1$ and a target state involving $N_B$ vibrons on mode $x=2$, with the cavity initially in the photon-free state. This choice was motivated by the fact that the AQD constitutes a prototype system that allows one to investigate the quantum counterpart of the well-known self-trapping mechanism.

Outside the cavity, initial and target states interact indirectly through their coupling to the remaining states of the $N_B$-vibron subspace. This interaction leads to the formation of two low-energy bound states in which the vibrons, tightly bound to each other, behave as a single particle delocalized over the two modes of the dimer. This delocalization is characterized by an effective hopping constant measured by the energy gap between the two bound states. Since the effective coupling decreases with the anharmonicity and the vibron number, the energy takes a very long time to tunnel from one mode to the other: this is the quantum manifestation of classical self-trapping.

When the dimer is placed inside the cavity, we have shown that the hybridization between the electromagnetic mode and the vibrational modes provides specific dynamical behaviors. The light-matter coupling increases the accessible Hilbert space, creating additional pathways between the initial state and the target state. Interference between these paths modifies the effective coupling that governs the transfer process. Depending on whether the interference is constructive or destructive, the effective hopping constant first decreases as $G$ increases from zero. The two lowest energy levels associated with the self-trapping dynamics approach one another. The energy gap decreases, which lengthens the time required for vibrational energy to migrate between the two modes. This effect can be interpreted as a light-induced self-trapping. Conversely, when the coupling increases, the trend reverses. This results in a rapid rise in level separation and an acceleration in energy transfer, leading to cavity-assisted energy transfer.
A key  feature is the existence of a critical parameter region where the gap between the two lowest states vanishes. In fact, for odd $N_B$ values, a particular light-matter coupling, $G_c$, generates an exact degeneracy. At the critical point, quantum interference almost suppresses the dynamics, resulting in the localization of most of the vibrational energy where it is deposited:  a stable QST occurs. This phenomenon is referred to as degeneracy-induced localization in quantum graph theory. It originates from the isomorphisms with vibron-photon dynamics and that of a single particle moving on a triangular network. 

In future work, it would be interesting to extend the present analysis to the characterization of several features that may  affect the energy transfer in the dimer. For instance, we have considered a cavity that is initially photon-free. A fundamental question is whether the initial photon number can influence the vibrational dynamics. As shown in Fig.~\ref{fig:networks}b, starting from a site other than the top of the triangle modifies the pathways to tunnel from one mode to another, thereby affecting the dynamics of the transfer. Similarly, the role played by the cavity frequency can be examined. By tuning this frequency, it should be possible to reshape the vibrational energy landscape and thereby open different pathways connecting the initial and target states. Finally, it would be worthwhile to go beyond the dimer and investigate whether this stabilized light-induced quantum self-trapping mechanism can also emerge in larger systems such as translationally invariant lattices.

\bibliography{mabiblio}

%----------------------------------------------------------------------------------------

\end{document}